\documentclass[slac_one]{revtex4}
\usepackage{graphicx}
\usepackage{fancyhdr}
\usepackage{amsmath}
\usepackage{amssymb}
\usepackage{epsfig}
\usepackage{euscript}
\begin{document}
\title{Towards Exact Quantum Loop Results in the Theory of General Relativity: Status and Update} 

%

\author{B.F.L. Ward}
\affiliation{Department of Physics,
 Baylor University, Waco, Texas, 76798-7316, USA}
%

\begin{abstract}
We present the status and update of  
a new approach to quantum general relativity as formulated by Feynman
from the Einstein-Hilbert action wherein 
amplitude-based resummation techniques are applied to
the theory's loop corrections to yield results (superficially)
free of ultraviolet(UV) divergences. Recent applications are summarized.
\end{abstract}

\maketitle

\thispagestyle{fancy}

The most basic law of physics, Newton's law, follows in a special case
of the classical solutions of Einstein's equation 
$$R^{\alpha\gamma}-\frac{1}{2}g^{\alpha\gamma}R+\Lambda g^{\alpha\gamma}  =-8\pi G_N T^{\alpha\gamma}$$
where $R=R^{\alpha}_{\alpha}$ is the curvature scalar, $R^{\alpha\gamma}$ is the contracted Riemann tensor, $T^{\alpha\gamma}$ is the energy momentum tensor, $g^{\alpha\gamma}$ is the metric of space-time, $G_N$ is Newton's constant and
$\Lambda$ is the cosmological constant. The many well-known
successful tests of the classical physics in Einstein's theory underscore
the need for an {\it experimentally verified} treatment of the quantum physics
in the theory. The most accepted approach is of course the superstring
theory~\cite{susystrg} and recently the loop quantum gravity 
formalism~\cite{lqg} has had success. Here we present the status and update
of a new approach which we have introduced in Refs.~\cite{bw1,bw2}
founded on amplitude-based resummation of the large 
infrared(IR) effects in the theory as
formulated by Feynman in Refs.~\cite{rpf1-2}. It does not 
modify Einstein's theory at all. We will see that it
makes contact with the phenomenological asymptotic safety fixed-point
approach in Refs.~\cite{reuter} as well. Our approach is thus seen to be
consistent with the second of the following four approaches to quantum 
gravity outlined in Ref.~\cite{wein1}: extension of the Einstein theory,
resummation, composite gravitons, and 
asymptotic safety -- fixed point theory.\footnote{The results in Refs.~\cite{qgrlarg}
on large distance effects in QGR are consistent with our
approach just as chiral perturbation theory in QCD is consistent
with perturbative QCD for short distance QCD effects.}
We thus proceed here as follows.
We start by reviewing briefly Feynman's formulation of Einstein's theory.
We follow this with a brief description of our resummed version of Feynman's
formulation. We close with the update of our recent applications.\par
More specifically, for the known world, we have the generally covariant 
Lagrangian
\begin{equation}
{\cal L}(x) = \frac{1}{2\kappa^2}\sqrt{-g} (R-2\Lambda)
            + \sqrt{-g} L^{\cal G}_{SM}(x)
\label{lgwrld1}
\end{equation}
where 
$g=\text{det}g_{\mu\nu}$, $\kappa=\sqrt{8\pi G_N}\equiv \sqrt{8\pi/M_{Pl}^2}$
where $M_{Pl}$ is the Planck mass,
and $L^{\cal G}_{SM}(x)$ is 
obtained from the usual Standard Model
Lagrangian $L_{SM}(x)$ by well-known general 
covariantization steps that are described
in Ref.~\cite{bw2}, for example. 
For reasons of pedagogy~\cite{rpf1-2}, we restrict our attention to 
the free massive physical Higgs scalar in $L_{SM}(x)$, $\varphi$, 
with a mass known to be greater than $114.4$ GeV with 95\% CL~\cite{lewwg}. 
Accordingly, we consider
the representative model~\cite{rpf1-2}
\begin{equation}
\begin{split}
{\cal L}(x) &= \frac{1}{2\kappa^2} R \sqrt{-g}
            + \frac{1}{2}\left(g^{\mu\nu}\partial_\mu\varphi\partial_\nu\varphi - m_o^2\varphi^2\right)\sqrt{-g}\\
            &= \quad \frac{1}{2}\left\{ h^{\mu\nu,\lambda}\bar h_{\mu\nu,\lambda} - 2\eta^{\mu\mu'}\eta^{\lambda\lambda'}\bar{h}_{\mu_\lambda,\lambda'}\eta^{\sigma\sigma'}\bar{h}_{\mu'\sigma,\sigma'} \right\}\\
            & \qquad + \frac{1}{2}\left\{\varphi_{,\mu}\varphi^{,\mu}-m_o^2\varphi^2 \right\} -\kappa {h}^{\mu\nu}\left[\overline{\varphi_{,\mu}\varphi_{,\nu}}+\frac{1}{2}m_o^2\varphi^2\eta_{\mu\nu}\right]\\
            & \quad - \kappa^2 \left[ \frac{1}{2}h_{\lambda\rho}\bar{h}^{\rho\lambda}\left( \varphi_{,\mu}\varphi^{,\mu} - m_o^2\varphi^2 \right) - 2\eta_{\rho\rho'}h^{\mu\rho}\bar{h}^{\rho'\nu}\varphi_{,\mu}\varphi_{,\nu}\right] + \cdots \\
\end{split}
\label{eq1-1}
\end{equation}
Here,
$\varphi(x)_{,\mu}\equiv \partial_\mu\varphi(x)$,
and $g_{\mu\nu}(x)=\eta_{\mu\nu}+2\kappa h_{\mu\nu}(x)$
where we follow Feynman and expand about Minkowski space
so that $\eta_{\mu\nu}=diag\{1,-1,-1,-1\}$.
We have introduced the notation~\cite{rpf1-2}
$\bar y_{\mu\nu}\equiv \frac{1}{2}\left(y_{\mu\nu}+y_{\nu\mu}-\eta_{\mu\nu}{y_\rho}^\rho\right)$ for any tensor $y_{\mu\nu}$\footnote{Our conventions for raising and lowering indices in the 
second line of (\ref{eq1-1}) are the same as those
in Ref.~\cite{rpf1-2}.}.
Thus, $m_o$ is the bare mass of our free Higgs field and we set the small
observed~\cite{cosm1} value of the cosmological constant $\Lambda$
to zero so that our quantum graviton has zero rest mass.
We return to this point, however, when we discuss phenomenology.
The Feynman rules for (\ref{eq1-1}) have been essentially worked 
out in Refs.~\cite{rpf1-2}, including the rule for the famous
Feynman-Faddeev-Popov~\cite{rpf1-2,ffp1} ghost contribution that must be added to
it to achieve a gauged-fixed unitary theory
(we use the gauge of Feynman in Ref.~\cite{rpf1-2},
$\partial^\mu \bar h_{\nu\mu}=0$),
so we do not repeat this
material here. We turn instead directly to the issue
of the effect of quantum loop corrections
in the theory in (\ref{eq1-1}).
\par
Specifically, the one-loop corrections to the graviton propagator
due to matter loops is just given by the diagrams in Fig. 1.
\begin{figure*}[t]
\centering
\includegraphics[width=85mm]{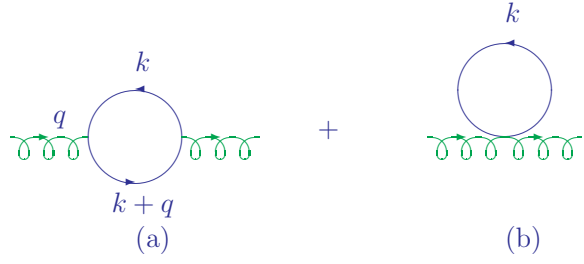}
\caption{\baselineskip=7mm     The scalar one-loop contribution to the
graviton propagator. $q$ is the 4-momentum of the graviton.} 
\label{fig1}
\end{figure*}
These graphs, with superficial degree of divergence 4, already illustrate the 
bad UV(ultra-violet) behavior of 
quantum gravity as formulated by Feynman. It is well-known
that theory is in fact non-renormalizable and we have proposed
amplitude-based exact resummation~\cite{bw1,bw2} as an approach
to deal with such bad UV behavior.
More precisely, from the electroweak resummed formula of Ref.~\cite{yfs} for the massive charged fermion proper self-energy for definiteness,  
\begin{equation}
\Sigma_F(p)=e^{\alpha B''_\gamma}\left[{\Sigma'}_F(p)-S^{-1}_F(p)\right]+
S^{-1}_F(p)
\end{equation}
which implies the exact result
\begin{equation}
iS'_F(p) = \frac{ie^{-\alpha B''_\gamma}}{S^{-1}_F(p)-{\Sigma'}_F(p)}
\label{yfsa}
\end{equation}
for
${\Sigma'}_F(p)=\sum_{n=1}^{\infty}{\Sigma'}_{Fn}$,
we need to find the quantum gravity analog of 
\begin{equation}
 \alpha B''_\gamma = \int d^4\ell\frac{S''(k,k,\ell)}{\ell^2-\lambda^2+i\epsilon}
\label{virt1}
\end{equation}
where $\lambda$ $\equiv$ IR cut-off and 
\begin{equation}
S''(k,k,\ell) = \frac{-i8\alpha}{(2\pi)^3}\frac{kk'}{(\ell^2-2\ell k+\Delta+i\epsilon)(\ell^2-2\ell k'+\Delta'+i\epsilon)}{\Big|}_{k=k'},
\end{equation}
$\Delta =k^2 - m^2$,~$\Delta' ={k'}^2 - m^2$. We show in 
Refs.~\cite{bw1,bw2} that the Feynman rules for (\ref{eq1-1})
lead to the results
\begin{equation} 
B''_g(k)= -2i\kappa^2k^4\frac{\int d^4\ell}{16\pi^4}\frac{1}{\ell^2-\lambda^2+i\epsilon}\frac{1}{(\ell^2+2\ell k+\Delta +i\epsilon)^2}
\label{yfs1} 
\end{equation}
and 
\begin{equation}
i\Delta'_F(k)|_{\text{resummed}} =  \frac{ie^{B''_g(k)}}{(k^2-m^2-\Sigma'_s+i\epsilon)}.
\label{resum}
\end{equation}
This latter equation is fundamental result. We stress already that, as
$\Sigma'_s$ starts in ${\cal O}(\kappa^2)$, we may drop it in
calculating one-loop effects and that explicit evaluation gives, 
for the deep UV regime,
\begin{equation}
B''_g(k) = \frac{\kappa^2|k^2|}{8\pi^2}\ln\left(\frac{m^2}{m^2+|k^2|}\right),
\label{deep}
\end{equation}
which shows that the resummed propagator falls faster than any power of $|k^2|$!(if $m$ vanishes, using the usual $-\mu^2$ normalization point we get\;
$B''_g(k)=\frac{\kappa^2|k^2|}{8\pi^2}
\ln\left(\frac{\mu^2}{|k^2|}\right)$
which again vanishes faster than any power of $|k^2|$! We show in Refs.~\cite{bw1,bw2} that these results render all quantum gravity loops finite.
We have called this representation 
of Einstein's theory resummed quantum gravity.\par
Turning now to the applications of our approach to quantum gravity, we note
that the respective resummed~\cite{bw1,bw2} prediction
for the graviton propagator implies~\cite{bw1,bw2} the
Newtonian potential
\begin{equation}
\Phi_{N}(r)= -\frac{G_NM}{r}(1-e^{-ar}),
\label{newtnrn}
\end{equation}
for
$a \cong  0.210 M_{Pl}$,
so that we agree with the phenomenological asymptotic safety approach of Refs.~\cite{reuter} for the UV fixed point behavior \begin{equation}
k^2G_N(k)=k^2G_N/(1+\frac{k^2}{a^2})\operatornamewithlimits{\longrightarrow}_{k^2\rightarrow\infty} a^2G_N=g_*\label{grun}\end{equation}
of the running Newton constant. Accordingly, we show in Refs.~\cite{bw1,bw2} that like Refs.~\cite{reuter} also find that elementary particles
with mass less than $M_{cr}\sim M_{Pl}$ have no horizons\footnote{See also Bojowald et al. in Refs.~\cite{lqg} for the analogous loop quantum gravity result.}. 
We have more recently shown
that the final state of Hawking radiation for an originally very massive
black hole is a Planck scale remnant with mass $\sim 2.4 M_{Pl}$ which may 
decay into cosmic rays that therefore 
have Planck scale energies~\cite{bw2}.
We have encouraged experimentalists to search for such.\par
In addition to our result for $g_*$ in (\ref{grun})
we also get UV fixed-point behavior for $\Lambda(k)/k^2$: using Einstein's equation
\begin{equation}
G_{\mu\nu}+\Lambda g_{\mu\nu}=-\kappa^2 T_{\mu\nu}
\end{equation}
and the point-splitting definition 
\begin{equation}
\varphi(0)\varphi(0)=\lim_{\epsilon\rightarrow 0}\varphi(\epsilon)\varphi(0)
=\lim_{\epsilon\rightarrow 0} T(\varphi(\epsilon)\varphi(0))
=\lim_{\epsilon\rightarrow 0}\{ :(\varphi(\epsilon)\varphi(0)): + <0|T(\varphi(\epsilon)\varphi(0))|0>\}
\end{equation}
we get for a scalar the contribution to $\Lambda$, in Euclidean representation,
\begin{equation}
\Lambda_s=-8\pi G_N\frac{\int d^4k}{2(2\pi)^4}\frac{(2\vec{k}^2+2m^2)e^{-\lambda_c(k^2/(2m^2))\ln(k^2/m^2+1)}}{k^2+m^2}
\cong -8\pi G_N[\frac{3}{G_N^{2}64\rho^2}],\;\;\;\text{for}\;\rho=\ln\frac{2}{\lambda_c}
\end{equation} 
with $\lambda_c=\frac{2m^2}{M_{Pl}^2}$.
For a Dirac fermion, we get~\cite{bw3} $-4$ times this contribution.
In this way, we get the UV limit, using $G_N(k)$ from (\ref{grun}),
\begin{equation}
\Lambda(k) \operatornamewithlimits{\longrightarrow}_{k^2\rightarrow\infty} k^2\lambda_*, \text{with}\;\;
\lambda_*\cong\frac{-1}{960\rho_{avg}}(\sum_jn_j)(\sum_{j}(-1)^{F_j}n_j)
\end{equation} 
where $F_j$ is the fermion number of $j$, $n_j$ is the effective
number of degrees of freedom of $j$, $1/\rho_{avg}$ is the attendant average
value of $1/\rho$ and we have used the result in eq.(17) in 
Ref.~\cite{bw3} for $a$ -- see also Refs.~\cite{bw1,bw2}.
It follows that all of the Planck scale cosmology 
results of Bonanno and Reuter~\cite{reuter} 
hold, but with definite results for the limits $k^2G_N(k)=g_*$ and
$\lambda_*$ for $k^2\rightarrow \infty$ -- we get $(g_*,\lambda_*)\cong (0.0442,0.232)$, to be compared with the estimates
in Refs.~\cite{reuter}, 
which give $(g_*,\lambda_*)\approx (0.27,0.36)$ 
and similar phenomenology~\cite{bw3}: we have 
a rigorous basis for solutions to the
horizon and flatness problems and the
scale free spectrum of primordial density fluctuations and initial entropy
problems by Planck and sub-Planck scale quantum physics.
We look forward to further applications of our approach 
to Feynman's formulation of Einstein's theory.

\begin{acknowledgments}
We thank Profs. S. Bethke and L. Stodolsky for the support and kind
hospitality of the MPI, Munich, while a part of this work was
completed. We thank Prof. S. Jadach for useful discussions.

Work partly supported
by the US Department of Energy grant DE-FG02-05ER41399
and by NATO Grant PST.CLG.980342.
\end{acknowledgments}

\end{document}